\documentclass[
  journal=pasa,
  manuscript=research-paper, 
  year=2022,
  volume=xx,
]{cup-journal}

\usepackage{siunitx,booktabs}
\usepackage{hyperref}
\usepackage{natbib}
\usepackage{amsmath}
\usepackage{amssymb}
\defcitealias{mardling2001}{MA01}
\defcitealias{vh22}{V+22}

\title{Empirical Stability Boundary for Hierarchical Triples}

\author{Max Tory}
\affiliation{School of Physics and Astronomy, Monash University, VIC 3800, Australia}

\author{Evgeni Grishin}
\affiliation{School of Physics and Astronomy, Monash University, VIC 3800, Australia}
\alsoaffiliation{OzGrav: Australian Research Council Centre of Excellence for Gravitational Wave Discovery, Clayton, VIC 3800, Australia}
\email[E. Grishin]{evgeni.grishin@monash.edu}

\author{Ilya Mandel}
\affiliation{School of Physics and Astronomy, Monash University, VIC 3800, Australia}
\alsoaffiliation{OzGrav: Australian Research Council Centre of Excellence for Gravitational Wave Discovery, Clayton, VIC 3800, Australia}

\doi{10.1017/pasa.2022.xx}

\received {dd Mmm YYYY}
\revised  {dd Mmm YYYY}
\accepted {dd Mmm YYYY}
\published{30 August 2022}


\begin{document}

\begin{abstract}
The three-body problem is famously chaotic, with no closed-form analytical solutions. However, hierarchical systems of three or more bodies can be stable over indefinite timescales. A system is considered hierarchical if the bodies can be divided into separate two-body orbits with distinct time- and length-scales, such that one orbit is only mildly affected by the gravitation of the other bodies. Previous work has mapped the stability of such systems at varying resolutions over a limited range of parameters, and attempts have been made to derive analytic and semi-analytic stability boundary fits to explain the observed phenomena. Certain regimes are understood relatively well.  However, there are large regions of the parameter space which remain un-mapped, and for which the stability boundary is poorly understood. We present a comprehensive numerical study of the stability boundary of hierarchical triples over a range of initial parameters. Specifically, we consider the mass ratio of the inner binary to the outer third body ($q_{\rm out}$), mutual inclination ($i$), initial mean anomaly and eccentricity of both the inner and outer binaries ($e_{\rm in}$ and $e_{\rm out}$ respectively). We fit the dependence of the stability boundary on $q_{\rm out}$ as a threshold on the ratio of the inner binary's semi-major axis to the outer binary's pericentre separation $a_{\rm in}/R_{\rm p, out} \leq 10^{-0.6 + 0.04q_{\rm out}}q_{\rm out}^{0.32+0.1q_{\rm out}}$ for coplanar prograde systems.  We develop an additional factor to account for mutual inclination. The resulting fit predicts the stability of $10^4$ orbits randomly initialised close to the stability boundary with $87.7\%$ accuracy.
\end{abstract}
\section{Introduction}
The three-body problem had been studied for over one hundred years \citep{poincare1899}, and is notoriously chaotic, with no analytic solution \citep{poincare1892}. For strongly chaotic systems a statistical approach is generally preferred to yield a probabilistic outcome of the scattering in terms of the final states \citep{moh1, moh2}. Only recently, the explicit dependence on the orbital elements was found with various methods \citep[e.g.][]{stone2019, ginat2, ginat1,kol2, kol1, Man2021}.
 
Hierarchical systems are a subset of three-body systems such that the system can be divided into an inner and an outer binary, each only weakly perturbed by the other. Hierarchical systems are deemed stable if they maintain their hierarchical structure without collision or ejection for a large number of orbits. The long-term evolution of hierarchical systems had been explored extensively over the years \citep{h68, ford2000, naoz13, toonen2020}.

Hierarchical systems are the only multiple systems observed in nature, either for triple stars of comparable masses \citep[where the majority of massive stars are actually in triples or higher multiples,][]{duchene13,moe17}, or systems with extreme mass ratios where one of the masses is much smaller than the other ones. Examples of such extreme mass ratio systems include (but are not limited to) circumbinary planets \citep{cbp1}, moons and binary asteroids \citep{rich06,np10}, or binary stars/black holes around an intermediate or super-massive black hole \citep{fg19}.

The boundary between stable hierarchical and unstable chaotic systems has a non-negligible width. For an inner separation $a_{\rm in}$, and an outer separation $a_{\rm out}$, the stability boundary (hereafter SB) will generally depend on the separation ratio $\alpha \equiv a_{\rm in}/a_{\rm out}$. It is expected that for $\alpha \sim 1$ the system is strongly chaotic, while it should be stable for $\alpha \ll 1$. The exact SB depends also on the masses of the bodies, as well as other orbital elements of both binaries. Due to the large parameter space and inherent chaotic nature of the problem, traditional studies explored the SB over a limited parameter space \citep{ek95, hw99, pet15, hp18, q18}, while new methods explore the boundary using a machine learning approach for specific choices of initial conditions (\citealp{lt22}; \citealp[][hereafter V+22]{vh22}). Of particular importance are the outer mass ratio $q_{\rm out} \equiv m_{\rm in}/m_{\rm out}$, and the mutual inclination between the two orbits, $i$. Here $m_{\rm in}$ is the total mass of the inner binary and $m_{\rm out}$ is the mass of the outer companion.

For extreme mass ratios $q_{\rm out}\ll1$, the Hill radius $r_{\rm H} \equiv a_{\rm out}\left(q_{\rm out}/3\right)^{1/3}\ll a_{\rm out}$ defines the length scale of the SB \citep{hill1878}. Tight inner binary orbits with $a_{\rm in} \ll r_{\rm H}$ will be stable while wide orbits with $a_{\rm in} \gtrsim r_{\rm H}$ will be unstable. Numerical studies have repeatedly shown that the  SB lies between between $0.4r_H-0.9r_H$ \citep{henon1970,hamilton1991}. The inclination dependence of the SB was examined analytically by \citet{innanen1979, innanen1980, myllari2018}, who explained the greater stability of retrograde orbits over prograde orbits by pointing out the asymmetric role of the Coriolis acceleration in stabilising the former and destabilising the latter. Secular analysis reveals more complicated behaviours such as the Lidov-Kozai (LK) mechanism in mutually inclined orbits, in which the outer perturber exerts a torque on the inner binary, exciting oscillations in inclination and eccentricity of the inner orbit \citep{lidov1962, kozai1962}. This reduces the critical stability radius for mutually inclined orbits.  \citet{grishin2017} generated a semi-analytic fit of the SB for arbitrary inclinations which explains the orbital distribution of irregular satellites of the giant planets \citep{carruba02}.

For comparable masses, $q_{\rm out} \lesssim 1$, instability (and later chaotic evolution and ejection) arises due to resonant interactions between the inner orbit and harmonics of the outer orbit \citep{mardling2008,mardling13}. Due to angular momentum exchange, the outer orbit gains eccentricity,  which induces higher harmonics. The structure of overlaps of neighbouring resonances leads to chaotic interactions \citep{chirikov79} and can be used to estimate stability. \citet[][hereafter MA01]{mardling2001} estimated the critical separation ratio for stability in terms of the mass ratio as $\alpha \propto (1+1/q_{\rm out})^{-2/5} \propto (q_{\rm out}/(1+q_{\rm out}))^{2/5}$. Although strictly valid for comparable masses, $q_{\rm out} \gtrsim 0.2$, numerous studies extrapolate the \citetalias{mardling2001} stability boundary to very high or very low mass ratios. Moreover, the \citetalias{mardling2001} inclination dependence is not particularly accurate. Very recently, \citetalias{vh22} updated the stability boundary to include non-monotonic inclination dependence.

Each empirical fit relies on an extensive parameter study and specific assumptions in the relevant regime which limit its domain of validity. However, unifying two or more regimes is challenging due to the very distinct underlying physical processes. For example, while the comparable masses case relies on exciting eccentricity in the outer orbit, the extreme mass ratio case cannot change the eccentricity of the outer orbit at all, and the instability is due to tidally shearing the inner binary. In addition, many of the recent population studies regularly check for violation of the SB during the dynamical evolution of a triple system \citep{toonen20,mse,gp22}. Finally, recent observations find companions with mass as low as $0.2M_\odot$ to massive O-stars, pushing the mass ratio to extreme values \citep{reggiani2022}. Given the large multiplicity of massive stars, it is plausible to find triples with extreme mass ratios. A unified stability boundary for wider range of mass ratios, inclinations and other parameters is compelling.

In this paper, we find a unified empirical fit for the stability boundary for any outer mass ratio, $q_{\rm out}$, and extend it to any mutual inclination. By doing so, we bridge together the $1/3$ power law for the extreme mass ratio regime and the $2/5$ regime for comparable masses. We also extend the inclination dependence fit and explore the dependence on the eccentricities and other orbital parameters. 
Our paper is organised as follows: Section 2 describes the numerical methods and initial conditions. Section 3 presents the results and the unified stability boundary fit, which is given in equations \ref{eqn:fit4_full_fit} through \ref{eqn:fit3_mratio_adjust}. In Section 4 we discuss the implications of our findings, and assess the utility of our fit against other contemporary models, and Section 5 summarises our findings.

\section{Methods}
All integrations were computed using the \href{https://rebound.readthedocs.io}{REBOUND} N-body code, with the IAS15 adaptive symplectic integrator, accurate to machine precision for a billion orbits \citep{ias15}. An orbit was considered disrupted if the distance between the inner binary components $m_1$ and $m_2$ exceeded half the outer pericentre separation $r_{\rm p, out} / 2 \equiv a_{\rm out} (1 - e_{\rm out}) / 2$. In orbits with $q_{\rm out} \le 1$, the inner binary can be disrupted before its separation exceeds $a_{\rm out}/2$. However in practice such orbits soon reach this separation, and the above criterion is sufficient to define stability, thus it is a conservative definition of stability. Reaching inner orbital separations of half the outer orbit’s pericentre also requires significant changes in the orbital energy, which coincides with other definitions of stability (e.g. \citetalias{vh22}). We compare different stability thresholds in sec. \ref{4.2}. Systems with large values of $q_{\rm out } \gg 1$ can be non-hierarchical and stable, where our criterion is not satisfied (\citealp{bhaskar21}, \citetalias{vh22}).


An orbit was classified as stable if it survived for 100 outer orbital periods ($P_{\rm out}$) without disruption, following previous work. \citet{grishin2017} integrated for $100 P_{\rm out}$ to incorporate the impact of secular changes that occur on timescales of $\lesssim 10 P_{\rm out}$ for systems close to the stability boundary. \citet{grishin2017} also ran one grid of initial conditions for longer times of $10^4 P_{\rm out}$ and obtained similar results. \citepalias{vh22} also used a simulation duration of $100 P_{\rm out}$, arguing that $99\%$ of the unstable orbits are classified as such within $100 P_{\rm out}$ (see their Fig. 1). However, the actual timescale for instability may be sensitive to initial conditions. While this paper was under review, \cite{hayashi22} integrated a limited sample of initial conditions for much longer timescales to show that some orbits (especially orbits with large $e_{\rm out}$ and large mutual inclinations) could reach instability much later, after $\gtrsim 10^4P_{\rm out}$. 

\subsection{Initialisation}
\begin{table}
    \centering
	\caption{Initial conditions for the plots presented in this paper. Variables represented by an interval are distributed uniformly unless given in exponential form (i.e. [$10^{-2}, 10^0$]), in which case they are distributed logarithmically. The value of $a_{\rm in}$ on the ordinate of each plot is scaled as appropriate to the figure.}
	\label{tab:init1}
	\begin{center}
    	\begin{tabular}{ | c | cccc | } 
    		\hline
    		\bf{Variable} & \bf{Fig. \ref{fig:1}} & \bf{Fig. \ref{fig:2}} & \bf{Fig. \ref{fig:3}} & \bf{Fig. \ref{fig:4}}\\
    		\hline
    		$i$ & [0,180] & 0, 150 & 0 & [0,180] \\
    		\hline
    		$e_{\rm in}$ & 0 & 0 & [0, 0.85] & 0 \\
    		\hline
    		$e_{\rm out}$ & 0 & 0 & [0, 0.85] & 0 \\
    		\hline
    		$q_{\rm in}$ & $10^{-2}$ & $10^{-2}$ & $10^{-2}$ & [$10^{-2}$, $10^{0}$] \\
    		\hline
    		$q_{\rm out}$ & 1, $10^{-5}$ & [$10^{-4}$, $10^0$] & $10^{-2}$ & $10^{-1}$, $10^{-3}$ \\
    		\hline
     		$\#$ of runs/$10^5$& $2\times4.68$ &  $2\times12$ &  $2\times1.36$ &  $4\times4.68$ \\
    		\hline
   
    	\end{tabular}
    \end{center}
\end{table}

\begin{figure*}[hbt!]
    \centering
    \includegraphics[width = 0.495\textwidth]{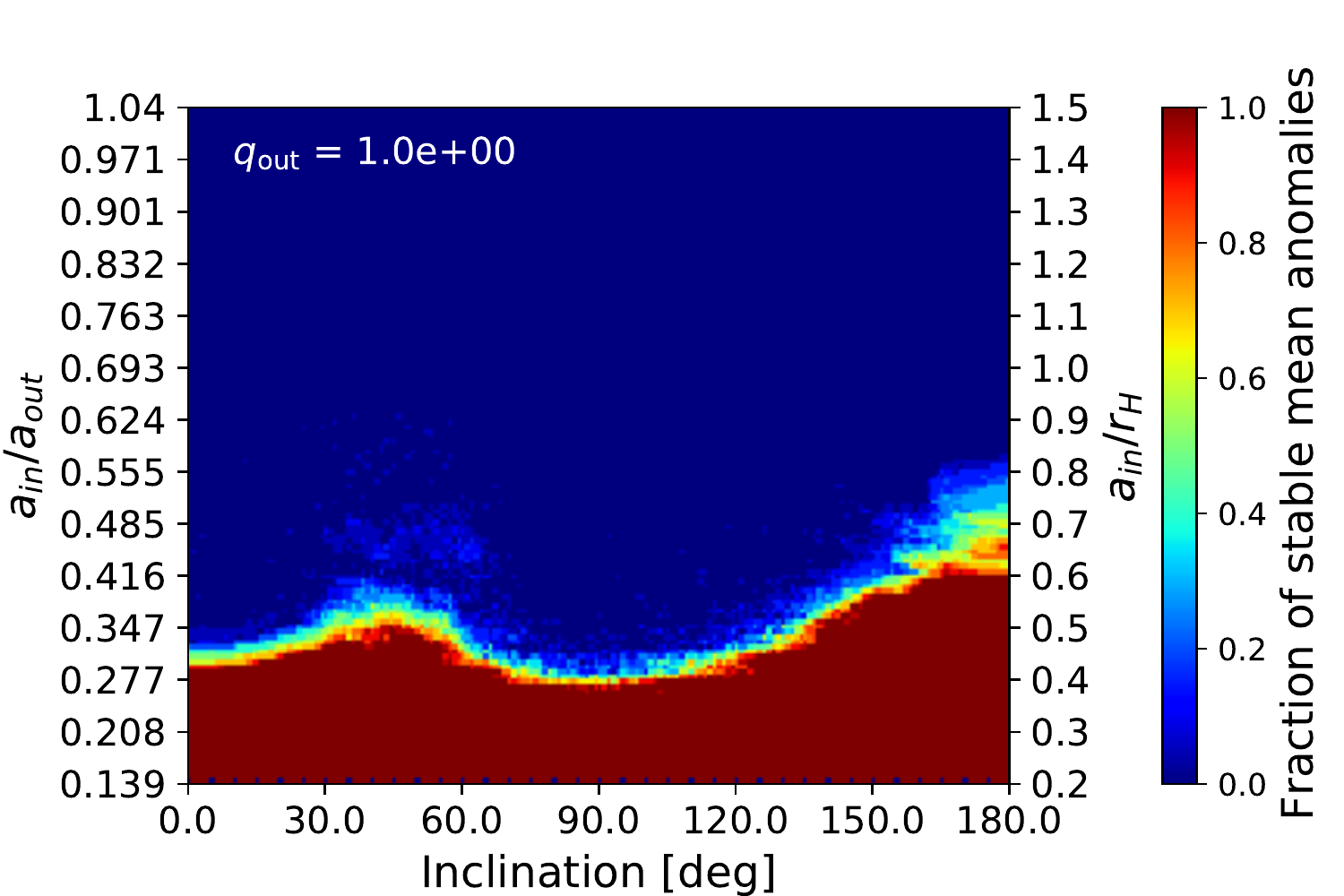}
    \includegraphics[width = 0.495\textwidth]{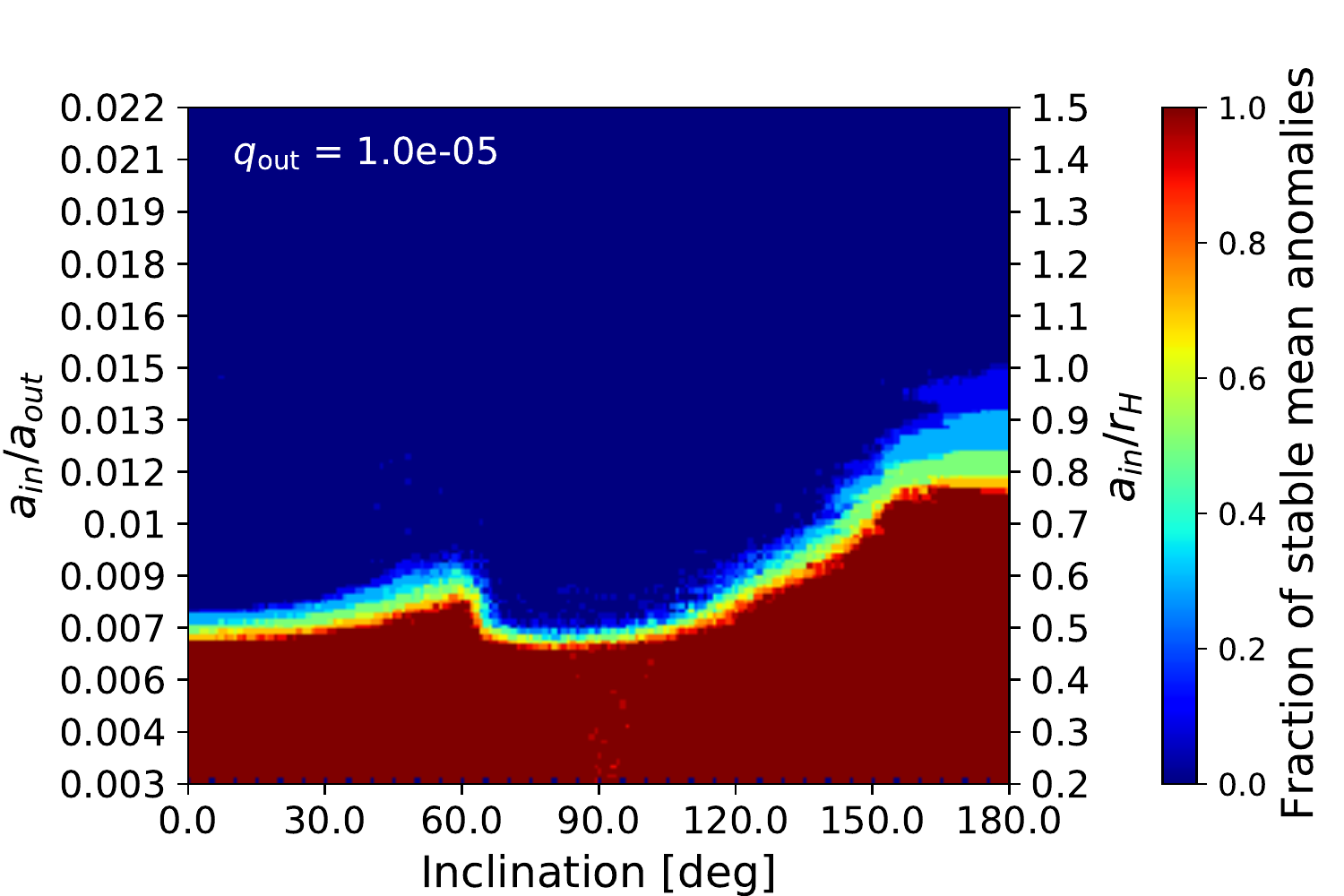}
    \caption{ Fraction of stable orbits out of 20 evenly spaced mean anomalies against mutual inclination for $q_{\rm out}=1$ (left) and $q_{\rm out}=10^{-5}$ (right). The left ordinate axis shows $a_{\rm in}$ in units of $a_{\rm out}$, while the right ordinate shows $a_{\rm in}$ in units of the Hill radius $r_{\rm H}$.} 
    \label{fig:1}
\end{figure*}

The system was initialised in Jacobi coordinates, described as follows. $m_1$ was initialised at the centre of the system initially with zero velocity. REBOUND initialisation tools were used to place $m_2$ in an orbit defined by $a_{\rm in}$ and $e_{\rm in}$. Finally, $m_{\rm out}$ was initialised in orbit around the centre-of-mass of $m_1$ and $m_2$, based on $a_{\rm out}$, $e_{\rm out}$ and inclination $i$. 

Each plot shows a high-resolution grid of systems, with one input parameter on the abscissa and a suitably scaled $a_{\rm in}$ on the ordinate. Eccentric orbits were initialised at apastron for greater precision. Since orbits at different mean anomalies exhibit slightly different stability, systems with 20 evenly-spaced mean anomalies were chosen for each grid coordinate, and the plots show the fraction of these systems that were stable at each coordinate. Red areas exhibited a high fraction of stable orbits, while the unstable regions are blue. For a detailed breakdown of initial conditions, see Table \ref{tab:init1}.

In order to obtain our results, we overall integrated over 5 million orbits. The last row of table \ref{tab:init1} lists the number of panels for each figure times the number of runs to generate each panel. 

\section{Results}
Here we explore the dependence of the SB on the various system parameters: the inclination (section 3.1), mass ratios (section 3.2) and eccentricity (section 3.3). We present our empirical fits in section 3.4. All logarithms in this paper are base 10.

\subsection{Inclinations} 

\begin{figure*}[hbt!]
    \centering
    \includegraphics[width = 0.49\textwidth]{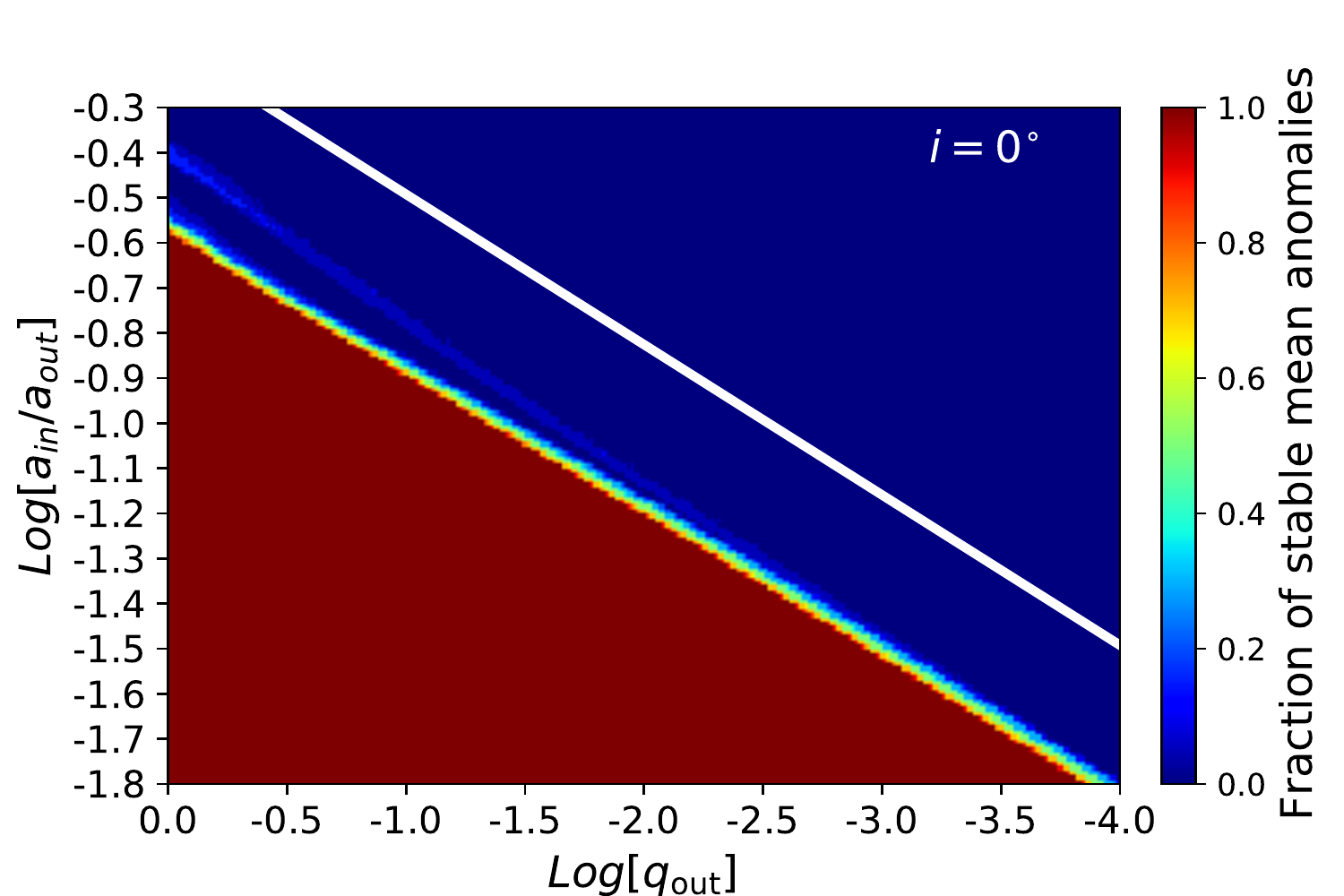}
    \includegraphics[width = 0.49 \textwidth]{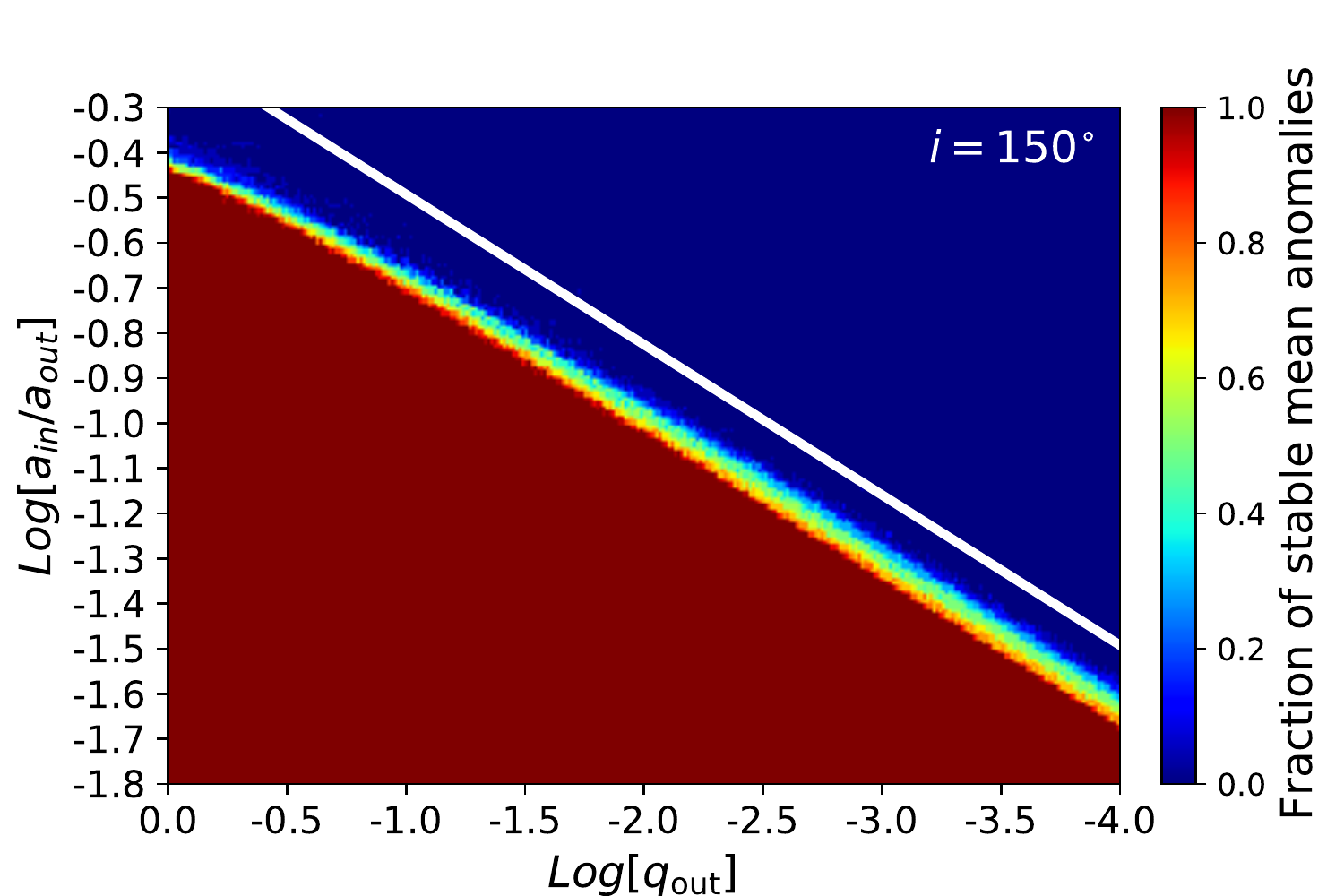}
    \caption{Fraction of stable orbits out of 20 evenly spaced mean anomalies against the mass ratio $q_{\rm out}$, for inclinations $i=0$ (left) and $i=150^\circ$ (right). The white line indicates the Hill radius $r_{\rm H}$ in both panels. At extreme mass ratios, the relationship is consistent with the Hill regime, while at $q_{\rm out} \geq 0.1$, high inclination orbits become relatively less stable, while low inclination orbits become more stable, although the inclination dependence is not monotonic (see Figure \ref{fig:1}).}
    \label{fig:2}
\end{figure*}

Fig. \ref{fig:1} shows the SB dependence on the mutual inclination. It closely follows previously reported non-monotonic fits \citep{hamilton1991, grishin2017}.  The boundary is essentially identical to the $q_{\rm out}\ll 1$ Hill approximation up to $\log[q_{\rm out}]\leq-1$, with prograde orbits stable up to around $0.5r_{\rm H}$, and retrograde orbits as far as $0.8r_{\rm H}$. Stability peaks at inclinations of $60^\circ$, and reaches maximal values again near $180^\circ$. The $\sim 90^\circ$ inclinations are the least stable, as expected from the prominent LK resonances that excite large eccentricities in this regime.

The left panel of Fig. \ref{fig:1} shows the $q_{\rm out}=1$ case. Here the prograde and retrograde SB's are similar with a small dip in the high inclination range. This is because the Hill approximation does not apply and instead the critical separation is somewhere between $\alpha=a_{\rm in}/a_{\rm out} \approx 0.3$ (for prograde) and $\alpha \approx 0.4$ -- $0.55$ (for retrograde), compatible with the \citetalias{mardling2001} and \citetalias{vh22} stability boundaries for both prograde and retrograde cases.

The SB also peaks around inclination $50^\circ$ more smoothly in the $q_{\rm out}=1$ case, while the decay is more abrupt in extreme $q_{\rm out}$ cases and occurs close to $60^\circ$. The increase of the critical inclination for the onset  of LK oscillations for triples with mild hierarchy (close to their stability limit) has been observed by \cite{grishin2017}, was later derived analytically by \cite{gpf18} for small $q_{\rm out}$, and was recently extended to any mass ratio by \cite{mg22}. These results are consistent with the analytic expectations. 

\subsection{Mass Ratios}

Figure \ref{fig:2} shows the dependence of the SB of the mass ratio $q_{\rm out}$. In the Hill approximation, where $\log q_{\rm out}<-1$, the power law of $1/3$ dominates across all inclinations. The inclination-dependent boundary of equal-mass systems manifests in a relative increase in the gradient of the SB towards $\log q_{\rm out} = 0$ in the low-inclination systems (left panel), and a corresponding decrease in the gradient of the stability boundary for more inclined systems (right panel).

\subsection{Inner and Outer Eccentricities}
\begin{figure*}[hbt!]
    \centering
    \includegraphics[width = 0.49\textwidth]{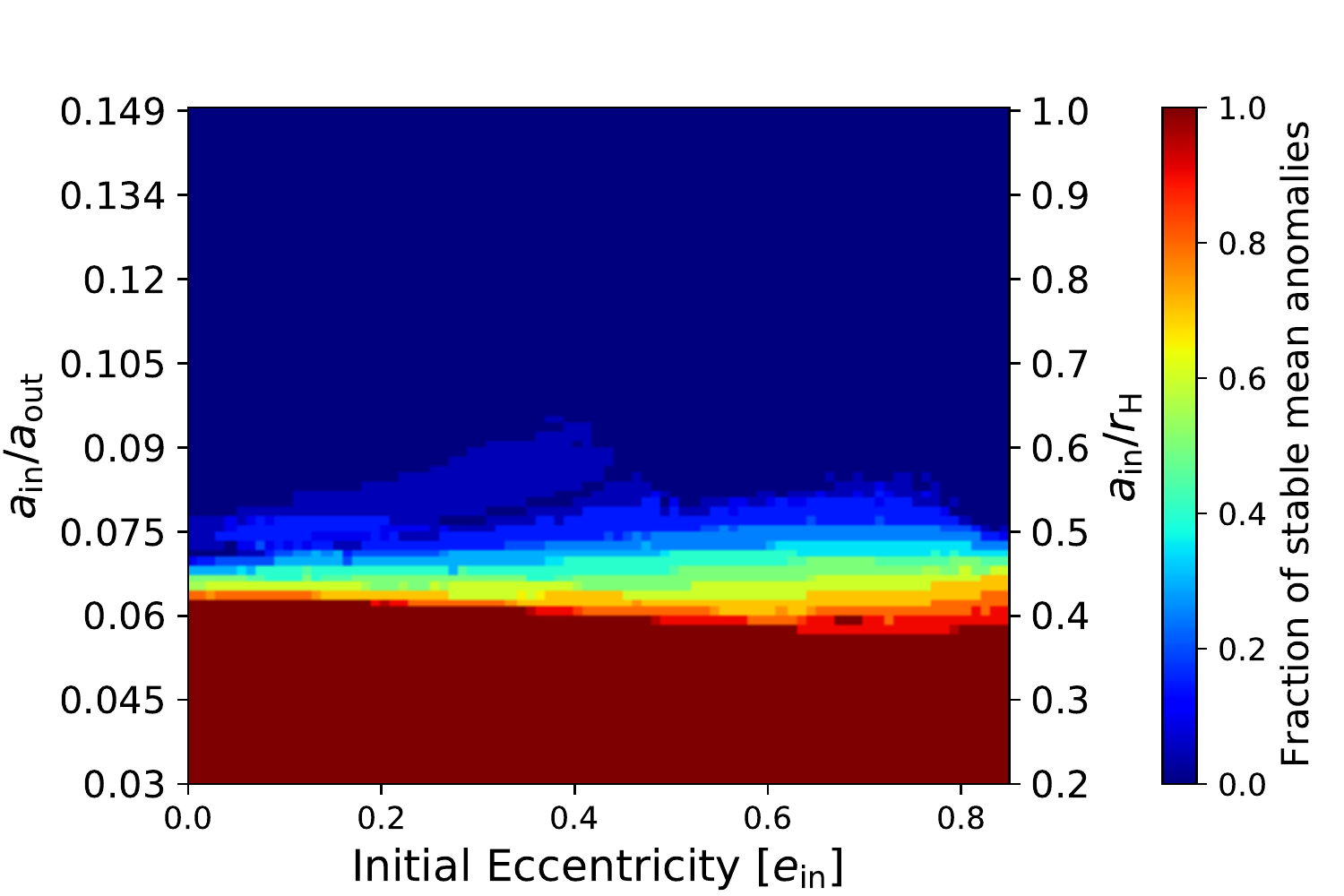}
    \includegraphics[width = 0.49\textwidth]{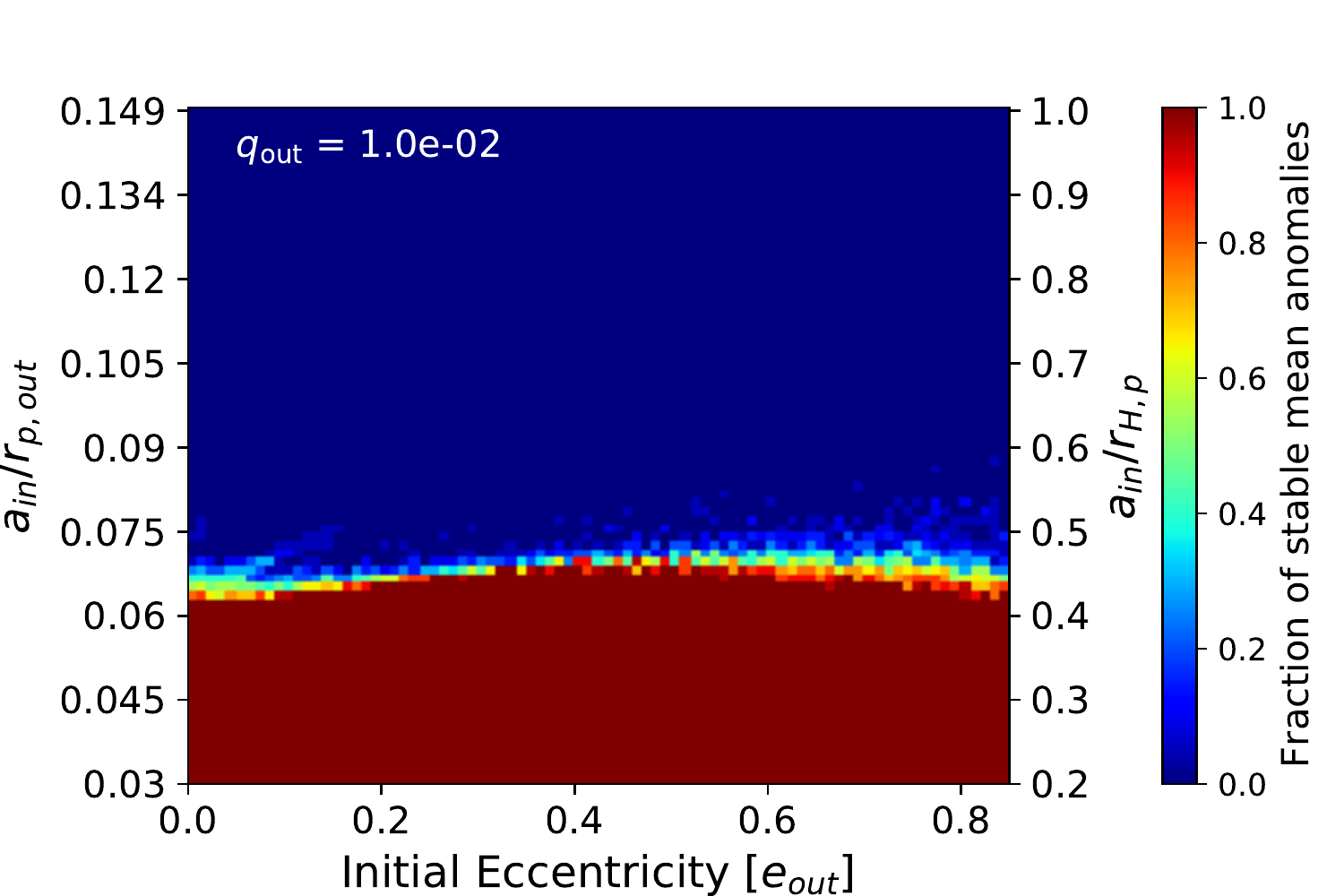}
    \caption{Fraction of stable orbits out of 20 evenly spaced mean anomalies against $e_{\rm in}$ (left) and $e_{\rm out}$ (right), plotted for $q_{\rm out}=0.01$. In the left panel, $e_{\rm out} = 0$, and in the right panel $e_{\rm in} = 0$.}
    \label{fig:3}
\end{figure*}

Previous work has demonstrated that in systems with $q_{\rm out} \sim 1$, the stability boundary of hierarchical triples depends on both outer and inner eccentricity \citepalias{mardling2001, vh22}. Preliminary consideration suggests that the circular stability boundary condition may be translated into one for a general eccentric boundary by considering the ratio between semi-major axis of the inner orbit $a_{\rm in}$ and the Hill pericentre radius $r_{\rm H,p} \equiv r_{\rm H} (1 - e_{\rm out})$, rather than $a_{\rm in}/r_{\rm H}$.  The impulse from the outer object is greatest at its pericentre. However, the eccentricity of the inner orbit is determined by the three-body interactions rather than by its initial value for $q_{\rm out} < 1$ (e.g., the maximum eccentricity in the Lidov-Kozai regime is set by the masses, separations and the mutual inclination \citealp{grishin2017,mg22}).  

Fig. \ref{fig:3} shows the dependence of the SB on the initial eccentricity, for $q_{\rm in}=q_{\rm out}=0.01$.  The left panel shows that the SB is largely independent of $e_{\rm in}$. The right panels shows that the dependence of the SB on the outer binary's eccentricity is indeed well approximated by using  $r_{\rm H,p}$ in place of $r_{\rm H}$ in the SB threshold.   Both of these approximations begin to break down when $q_{\rm out} \gtrsim1$.  The initial eccentricity of the inner binary is more significant when $q_{\rm out} \gtrsim 1$; in this case, the apocentre separation of the inner orbit becomes a relevant parameter $r_{\rm a, in} \equiv a_{\rm in} (1+e_{\rm in})$, since $m_2$ is most weakly attracted to $m_1$ at apocentre (see \citepalias{mardling2001, vh22} for a discussion of eccentricity in this regime).

\subsection{Fits}

Constructing a fit to a multivariate parameter space entails significant challenges, due to the numerous potential correlations between the various parameters. In order to tackle this problem, we first attempt to approximately describe the SB by the product of single-variable functions for each parameter, since the plots for inclination look similar at different mass ratios. 
Furthermore, since $e_{\rm in}$ is mostly irrelevant and the effects of $e_{\rm out}$ can be explained by considering the critical separation as a fraction of the outer periastron separation, the problem is reduced to only two variables, inclination and mass ratio: 
\begin{equation}
    \frac{a_{\rm in}}{R_{\rm p, out}} = f(q_{\rm out})\cdot g(i) \cdot h(q_{\rm out}, i).
    \label{eqn:fit4_full_fit}
\end{equation}

A piecewise linear regression with two segments was used to describe the log-log mass ratio boundary accurately in the low-inclination limit, and a smooth interpolation between these two regimes provides the first factor, $f(q_{\rm out})$:
\begin{equation}
    f(q_{\rm out}) = 10^{-0.6 + 0.04q_{\rm out}}q_{\rm out}^{0.32+0.1q_{\rm out}},
    \label{eqn:fit1_mratio}
\end{equation}
where the numerical pre-factor is chosen to optimise classification accuracy across eccentricities (see section 4 for more details on classification accuracy).

The inclination boundary is more complicated, and requires more careful treatment. A cosine function describes the curve up to $60^\circ$, while the high-inclination region is fit by a fourth-order polynomial. The combination of these equations gives the function $g(i)$ (here, all inclinations are given in radians): 
\begin{align}
    & g(i) = \\ 
    & \Bigg\{    \begin{array}{ll}
            -0.4 \cos{i} + 1.4 & \ \ i < \pi / 3\\
            -0.1773 i^4 + 1.1211 i^3  - 1.9149 i^2 +0.5022 i + 1.6222 & \ \ i \geq \pi/3
        \end{array} \nonumber
    \label{eqn:fit2_inc_poly}
\end{align}

Given these two independent fits, it is necessary to account for the mutual interaction between inclination and mass ratio. The factor $h(q_{\rm out}, i)$ reduces stability at high inclinations for roughly equal-mass systems (Eq. \ref{eqn:fit3_mratio_adjust}). Thus, the empirical fit to the stability is given by equation \ref{eqn:fit4_full_fit} with 

\begin{equation}
    \log h(q_{\rm out},i) = \frac{-iq_{\rm out}^{1.3}}{1500} \ .
    \label{eqn:fit3_mratio_adjust}
\end{equation}

\section{Comparison to other stability boundary fits}

\begin{figure*}[hbt!]
    \centering
    \includegraphics[width = \textwidth]{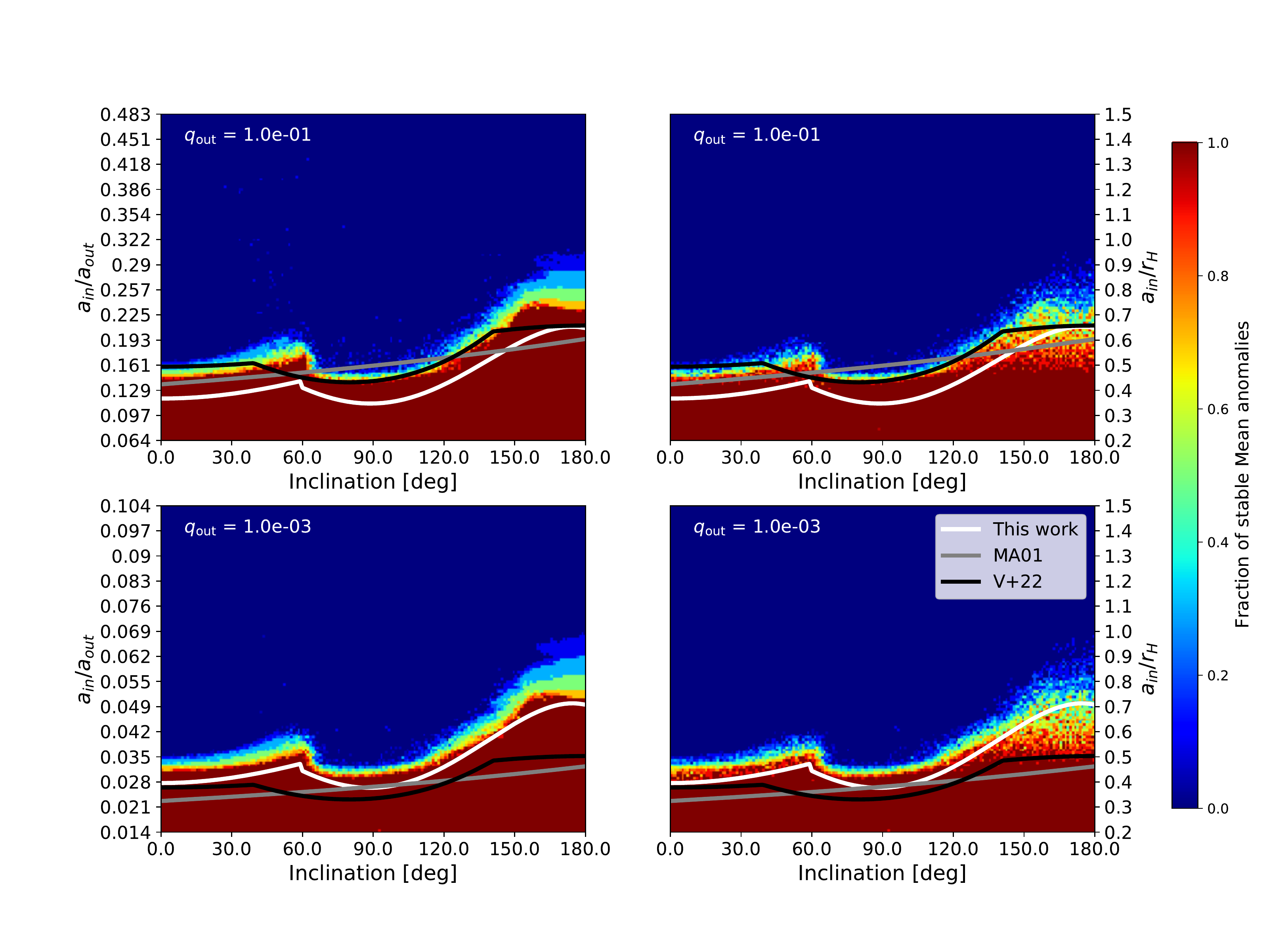}
    \caption{Fraction of stable orbits out of 20 evenly spaced mean anomalies against $i$, plotted for $q_{\rm out}=0.1$ (top row) and $0.001$  (bottom row). Left panels: our threshold for instability ($a_{\rm in}>a_{\rm out}/2$). Right panels: \citetalias{vh22}'s threshold for instability (either separation varies by $>10\%$). }
    \label{fig:4}
\end{figure*}

Here we compare our stability boundary with other works. We first summarise the fractions of orbits correctly classified as stable/unstable, then directly compare to the recent algebraic fit from \citetalias{vh22} and discuss the performance of different instability thresholds and their regions of validity. 

\subsection{Overall fractions of correctly classified orbits}
Equation \ref{eqn:fit4_full_fit} was compared to other well-known hierarchical stability boundary models by predicting stability for $10^4$ orbits initialised according to Table \ref{tab:init2}. In order to focus on the regime of interest and better differentiate between the quality of the different fits, $a_{\rm in}/R_{\rm p, out}$ was restricted to be between 0.5 and 1.5 times the stability boundary as described by equation \ref{eqn:fit4_full_fit}. This results in less impressive classification scores than are found in other work \citepalias{vh22}, since we focus on the region in which it is most difficult to predict stability. As in Section 2, we classified an orbit as stable if the distance between $m_1$ and $m_2$ remained smaller than $r_{\rm p, out}/2$ for 100 outer orbital periods. Equation \ref{eqn:fit4_full_fit} predicted stability correctly for 87.7\% of orbits, compared to 67.4\% for the fit proposed by \citetalias{mardling2001} and 70.5\% for the adjusted fit put forward in \citetalias{vh22}. Of the misclassified orbits, 53.3\% were unstable (classified stable) and the rest were stable, but classified as unstable, compared to 97.6\% and 98.6\% of misclassified orbits being stable (classified as unstable) for the \citetalias{mardling2001} and \citetalias{vh22} fits respectively. This stark discrepancy arises from the inconsistency of the 2/5 power law with the 1/3 power law Hill regime in mass-hierarchical systems, in which both the \citetalias{mardling2001} and \citetalias{vh22} fits drop off too quickly with $q_{\rm out}$, causing them to over-classify orbits as unstable. 

\subsection{Varying the instability threshold} \label{4.2}
The definition of stability used in this paper differs somewhat from that put forward by \citetalias{vh22}, who extend instability to include any system for which $a_{\rm in}$ or $a_{\rm out}$ change by more than 10\% after 100 $P_{\rm out}$, in order to catch systems which have not yet been disrupted but will be in the near future. Figure \ref{fig:4} compares the stability boundaries put forward by this work, \citetalias{mardling2001} and \citetalias{vh22} against inclination, for different mass ratios and different stability criteria.

It is clear that equation \ref{eqn:fit4_full_fit} most closely approximates the stability boundary when $q_{\rm out} \ll 1$, due to its adherence to the Hill mass-ratio dependence in this regime. Since the \citetalias{mardling2001} and \citetalias{vh22} stability boundaries both approach a 2/5 power law at small $q_{\rm out}$, they underestimate the maximum stable $a_{\rm in}$ in these systems. At $q_{\rm out} = 0.1$, the stability boundary is best fit by the \citetalias{vh22} formula.

As for the different definitions of stability used, they seem to be largely consistent below $\sim120^{\circ}$ inclination, however in retrograde systems, the \citetalias{vh22} definition provides a much broader region of ambiguous stability, with a fraction of orbits being labeled stable depending on the initial mean anomaly. This is probably due to the large chaotic extent of meta-stable orbits that are unstable according to the \citetalias{vh22} criterion, but are stable for our criterion. This retrograde region is accurately fit by the \citetalias{vh22} stability boundary in regimes of moderate $q_{\rm out}$, however that fit underestimates the high inclination stability boundary when using our definition. 

For $10^{-6} < q_{\rm out} < 1$,  our definition of stability appears to provide a more clearly defined, smooth, boundary. However, for $q_{\rm out} > 1$, the \citetalias{vh22} definition is more appropriate, as it considers changes to both $a_{\rm in}$ and $a_{\rm out}$. In these systems, instability can mean ejection or disruption of $m_{\rm out}$, which may not be detected by the definition of stability we use.

The stability boundary described by Equation \ref{eqn:fit4_full_fit} unifies our previous understandings of the Hill regime \citep{grishin2017} and the similar-mass regime \citepalias{mardling2001}. It provides an accurate prediction of stability for any system with $q_{\rm out} \lesssim 1$, and captures the complexity of the inclination dependence. However, the dependence of the $a_{\rm in}/a_{\rm out}$ SB cannot be extrapolated far beyond $q_{\rm out} = 1$, since $a_{\rm in}$ and $a_{\rm out}$ become comparable at the SB, and the system ceases to be hierarchical. Hence, in that regime it is necessary to adopt a fit that asymptotes toward independence from $q_{\rm out}$ \citepalias{vh22, mardling2001}.

\begin{table}
    \centering
	\caption{Range of orbital elements used for fit evaluation. Inclination followed an isotropic distribution, mass ratios were distributed logarithmically, and all other elements were distributed uniformly. $a_{\rm in, crit}$ refers to the critical $a_{\rm in}$ of the stability boundary predicted by Equation \ref{eqn:fit4_full_fit}.}
	\label{tab:init2}
	\begin{center}
    	\begin{tabular}{ | cccccc | } 
    		\hline
    		\bf{$a_{\rm in}/a_{\rm in, crit}$} & \bf{$\log q_{\rm in}$} & \bf{$\log q_{\rm out}$} & \bf{$i$} & \bf{$e_{\rm in}$} & \bf{$e_{\rm out}$}\\
    		\hline
    		[0.5,1.5] & [-2,0] & [-6,0] & [0,$\pi$] & [0,1] & [0,1]  \\
    		\hline
    	\end{tabular}
    \end{center}
\end{table}

\section{Conclusions}

We have obtained an empirical fit to hierarchical three-body stability for systems with $q_{\rm out} \lesssim 1$. This fit attempts to bridge the gap between the well understood hierarchical-in-mass Hill regime \citep{grishin2017} and previous understandings of the equal-mass regime \citepalias{mardling2001, vh22}. We have mapped the stability boundary extensively across mass ratio and inclination within this range, and determined a 'turn-off' point from the Hill regime at $q_{\rm out} \approx 0.1$. We have also confirmed the accuracy of our fits through an extensive test, correctly predicting stability in 87.7\% of systems on or near the SB.

Our empirical fit captures the features of the non-monotonic inclination dependence of the stability boundary more accurately than other contemporary works which derive semi-analytic fits \citep{mardling2001, grishin2017, vh22}. Furthermore, it is the only stability boundary fit which unifies the resonant equal-mass regime and the Hill regime, dominated by tidal shearing. This is reflected in its classification performance, where it out-performs alternative algebraic stability boundary models for systems with $q_{\rm out} \ll 1$ while maintaining a comparable accuracy for those with $q_{\rm out} \approx 1$. 

The stability boundary fit presented in this paper is limited by its treatment of $q_{\rm out}$ as it approaches 1. Equation \ref{eqn:fit1_mratio} defines a stability boundary which increases indefinitely with $q_{\rm out}$, and with an increasing gradient. This arises from a somewhat myopic analysis of the behaviour of $q_{\rm out}$ in co-planar prograde systems, a relationship which is not borne out in more inclined systems (see Fig.~\ref{fig:2}), and cannot be sustained as $q_{\rm out}$ exceeds 1. Furthermore, this work deals only in passing with eccentricity as a factor impacting stability, and does not attempt to consider its interaction with inclination, nor to model its effect beyond a simple linear fit. This approximation is accurate for $q_{\rm out} < 0.1$, but fails in the equal-mass regime and beyond. 

Equation \ref{eqn:fit4_full_fit} provides a promising foundation for a generalised stability boundary for all hierarchical triples. Future work should consider the fits for systems with $q_{\rm out} > 1$ provided by \citetalias{mardling2001}, \citetalias{vh22} and others, and expand our study of the parameter space to evaluate the interactions of eccentricities with other parameters, in order to construct this unified fit. Analytic future work may also reveal the exact transition between the resonant and Hill regimes, to be compared with our empirical findings.

The study of populations of triples and their evolution can be numerical \citep{anderson17, toonen20,mse,gp22}, or semi-analytical \citep{munoz16,mg22}. However, even numerical population studies generally require secular approximations to reduce computational costs, and therefore rely on dynamical stability checks during the evolution.  Our new stability boundary fit will aid in distinguishing stable and unstable systems, which will most likely affect the overall branching ratios of the different outcomes.

Recently, machine learning tools have also been utilised to study triple stability \citep{lt22,vh22}. Due to their complexity, large neural networks have the potential to capture non-linear multi-dimensional interactions between parameters more accurately than any algebraic fit to the stability boundary, and further study should endeavour to apply these methods to a wider range of systems for more broadly applicable classification. However, neural networks could be harder to implement in simple population-synthesis style studies, where our simple algebraic stability fit will be useful. 

\section*{Acknowledgements}

We thank Pavan Vynatheya for useful discussions, Rosemary Mardling for helpful discussions and comments on the manuscript, and the anonymous referee for useful feedback. E.G.~and I.M.~acknowledge support from the Australian Research Council Centre of Excellence for Gravitational  Wave  Discovery  (OzGrav), through project number CE17010004. I.M.~is a recipient of the Australian Research Council Future Fellowship FT190100574.  Part of this work was performed at the Aspen Center for Physics, which is supported by National Science Foundation grant PHY-1607611.  The participation of I.M.~at the Aspen Center for Physics was partially supported by the Simons Foundation. This work made use of the OzSTAR high performance computer at Swinburne University of Technology. OzSTAR is funded by Swinburne University of Technology and the
National Collaborative Research Infrastructure Strategy (NCRIS).




\appendix

\end{document}